%% file: ICRC2025_template_IceCube.tex
\documentclass[a4paper,11pt]{article}

\usepackage{pos}
\usepackage{lipsum} 
\usepackage{ragged2e}
\title{Multi-Energy and Multi-Sample Searches for IceCube Neutrinos from LIGO/Virgo/KAGRA Gravitational Wave Events}

\ShortTitle{Multi-Energy and Multi-Sample Searches for Neutrinos from  GW Events}

\author{The IceCube Collaboration \\{\normalsize \normalfont(a complete list of authors can be found at the end of the proceedings)}\\}

\emailAdd{abalagopalv@icecube.wisc.edu}
\emailAdd{shori@icecube.wisc.edu}
\emailAdd{jvandenbroucke@icecube.wisc.edu}

\abstract{

The IceCube Neutrino Observatory at the South Pole detects neutrinos of astrophysical origin via their interactions with ice. The main array is optimized for the detection of neutrinos with energies above 1 TeV. A much smaller infill array, known as IceCube DeepCore, extends the sensitivity down to a few GeV. Neutrinos observed in both parts of the detector are used for astrophysical-source searches with multiple messengers. We present two analyses that follow up archival gravitational wave (GW) events from runs O1 through O3 of LIGO/Virgo/KAGRA. The first analysis uses two neutrino datasets: one with high-energy tracks and another consisting of low-energy tracks and cascades. These two neutrino datasets were previously used independently to follow-up GW events. In the analysis presented here, a combined likelihood search is performed using both datasets to search for neutrinos coincident with the GW events across a wide energy range, from a few GeV to several PeV. The second analysis, for the first time, uses a neutrino-induced cascade sample with events of energy above ~1 TeV for searches of coincident neutrino-GW emission. We present results from both analyses and discuss prospects for conducting these analyses in real time.

\vspace{4mm}

{\bfseries Corresponding authors:}
Aswathi Balagopal V.$^{1*}$, 
Sam Hori$^{2}$, 
Justin Vandenbroucke$^{2}$\\
{$^{1}$ \itshape University of Delaware}\\
{$^{2}$ \itshape University of Wisconsin-Madison}
\\[4mm]
$^*$ Presenter
}

\input{ICRCdetails.tex}

\begin{document}

\maketitle

\section{Introduction}\label{sec1}
Mergers of binary black holes (BBH), binary neutrino stars (BNS), or a neutron star and a black hole (NSBH) are seen to produce gravitational waves (GW) detectable by the LIGO/Virgo/Kagra (LVK) instruments. Such events are known to produce multiple messengers, as observed during the first confirmed BNS GW170817, known to have produced a gamma-ray burst and photons across the electromagnetic spectrum. These binary mergers can also be potential sites of neutrino production, in particular, if the merger involved a neutron star. Neutrino production can occur either in short time scales, within a 1000 s as motivated by~\cite{Baret:2011tk} or on longer time scales for binaries with at least one neutron star \citep{Fang:2017tla}.

The IceCube Neutrino Observatory, at the South Pole, detects neutrinos with energies of hundreds of GeV and above~\cite{Aartsen:2016nxy}. IceCube has been used to search for high energy (TeV to PeV) muon neutrinos from binary mergers producing GWs detected by LVK~\cite{2020ApJ...898L..10A,IceCube:2022mma}. These searches have been done both in real time, when alerts were issued publicly by LVK since its O3 run, and offline based on the candidate events published in the catalogs GWTC-1~\cite{2019PhRvX...9c1040A}, GWTC-2.1~\cite{2021arXiv210801045T} and GWTC-3~\cite{2021arXiv211103606T}. The real time search is also ongoing for the current O4 run of LVK. Archival searches were also performed on GW events from these catalogs using low energy (sub-TeV) neutrinos observed by IceCube. These neutrinos were detected by IceCube's denser infill array, which features smaller spacing between optical modules and enables the detection of sub-TeV events~\cite{IceCube:2023atb}.  Additionally, IceCube has searched for MeV - and GeV-scale~\cite{2021arXiv210513160A} neutrinos from these GW events.

Here we present two new analyses with IceCube neutrinos correlated to GW events detected by LVK from O1 to O3, published in the catalogs GWTC-1, GWTC-2.1 and GWTC-3. The first analysis searches for cascade neutrinos detected by IceCube correlated to the GW events. The second analysis performs a joint fit with low and high energy neutrino datasets, previously used for individual follow-up searches of GW events.
\section{Event Selections}\label{sec2}
\subsection{IceCube: Detector and Events}
IceCube, consisting of 86 strings hosting 5160 optical modules, detects neutrinos via the Cherenkov light emitted by their secondaries as they interact with ice~\cite{Aartsen:2016nxy}. The detector's main array is optimized for neutrinos with TeV to PeV energies. IceCube also has an infill array of 7 strings with closer spacing between the modules, that allows for a lowering of the energy threshold~\cite{IceCube:2011ucd}.
This array, known as IceCube DeepCore, allows the detection of neutrinos with enegries of a few GeV and above. IceCube observes two main types of event classes: cascades and tracks. Cascades are generated by charged-current (CC) interactions of electron neutrinos and neutral current (NC) interactions of neutrinos of all flavor.
Tracks are generated as a result of CC interactions of muon neutrinos.
The long lever arm of track events allow for a good angular resolution, while the contained nature of most cascades result in a good energy resolution.
Tracks in the southern sky have a huge background of muons from cosmic-ray air showers, and therefore track samples are harshly cut in the southern sky, affecting their sensitivities. Cascades, by nature of their event morphology, are easily separated from atmospheric muons, and therefore provide better sensitivity in the southern sky than tracks.

\subsection{Cascades sample}

The high-energy cascade analysis uses a machine learning based event selection which provides a large improvement over previous IceCube cascade samples. This sample provided the first evidence of the Galactic Plane as a source of neutrinos \cite{IceCube_GalacticPlane}. The sample has a sensitive energy range of $\sim 500\,\mathrm{GeV}$ to $1\,\mathrm{PeV}$. The cascade dataset has a low rate ($\sim0.2$\,mHz compared to $\sim6$\,mHz for tracks) with excellent effective area and purity across the entire sky.  The improved effective area of this sample in the southern sky when compared to tracks is an advantage while searching for neutrino events coincident with GW events. Although IceCube cascade events are more difficult to reconstruct than tracks, the larger uncertainty on the origin of the neutrinos is less important for transient analysis due to the reduced background.

\subsection{High energy tracks sample and low energy cascades and tracks sample}
The second analysis uses two datasets: a high-energy tracks sample, called GFU~\cite{Kintscher:2016uqh}, and a low energy sample consisting of both tracks and cascades, called GRECO Astronomy~\cite{IceCube:2022lnv}. The GFU sample primarily consists of muon-neutrino induced tracks, with energies of hundreds of GeV and above in the northern hemisphere, and several tens of TeV and above in the southern hemisphere. The GRECO Astronomy dataset contains all-flavor neutrinos with interaction vertices inside IceCube DeepCore, spanning energies from a few GeV to a few tens of TeV. These datasets are complimentary in their sensitivities and allows for a broad coverage in energy ranges of the neutrino events. Therefore, it is suitable to use them in a joint fit to leverage their respective strengths and search for neutrinos correlated to GW events across a wide energy range.
\section{Analysis Method}\label{sec3}
We search for neutrinos correlated to GW events within a 1000-second time window with respect to the time stamp of the GW event. The cascades analysis additionally searches for neutrinos within a 2-week period for GW events that include a neutron star. This search is not done for the combined analysis since the GRECO Astronomy dataset performs weakly for long time-window searches. Both analyses use an unbinned maximum likelihood (UML) method to perform the search. 
We define a likelihood of the form
\begin{equation}
 \mathcal{L} = \frac{(n_{\mathrm{s}}+n_{\mathrm{b}})^{N}}{N!}e^{-(n_{\mathrm{s}}+n_{\mathrm{b}})}\,\prod_{i=1}^{N} \left(\frac{n_\mathrm{s}\mathcal{S}_{i}}{n_{\mathrm{s}}+n_{\mathrm{b}}} + \frac{n_\mathrm{b} \mathcal{B}_{i}}{n_{\mathrm{s}}+n_{\mathrm{b}}}\right),
 \label{eq:likelihood}
\end{equation}
where $n_{\mathrm{s}}$ and $n_{\mathrm{b}}$ represent the signal and background neutrinos, $\mathcal{S}_{i}$ and $\mathcal{B}_{i}$
 the signal and background PDFs defined as in~\cite{Braun:2008bg}. This likelihood is maximized on all pixels in the sky, with an additional penalty term derived from the GW skymap applied to it. This gives the test statistic (TS) of the analyses. This TS is used to obtain sensitivities of the analyses, and finally determine the significance of our observed data when compared to background. The basic methodology follows the same procedure as shown in~\citep{2020ApJ...898L..10A, IceCube:2022mma,IceCube:2023atb}. While the intial UML follow-up searches with IceCube also fit the spectral index ($\gamma$)~\cite{IceCube:2022mma}, the two analyses reported here fix $\gamma$ to 2.5, following the shape of the diffuse flux of astrophysical neutrinos detected by IceCube~\cite{IceCube:2024fxo}.
 \subsection{Method for Combined GRECO \& GFU analysis}
 While the cascades analysis performs the fit only with one IceCube dataset, the combined analysis performs a joint fit using the high energy tracks and low energy all-flavor datasets described in section~\ref{sec2}. 
 To ensure dataset independence and peform a joint fit, events appearing in both datasets are removed from GRECO Astronomy and retained in GFU, which demonstrates a better reconstruction. This prevents double counting and allows for a consistent joint analysis. As these overlapping events represent only 1\% of the data, the Monte Carlo simulations are also treated as independent without any further processing. 

 The log-likelihood is evaluated individually for both the GRECO and GFU datasets at each pixel, and then added to obtain the combined likelihood. Since the spectral index is fixed, this simply amounts to a scanning of the likelihood in $n_\mathrm{s}$ for both datasets, adding them up, and evaluating the combination that maximizes the likelihood.
 \begin{equation}
     \ln\mathcal{L}_{\mathrm{combined}} = \ln\mathcal{L}_{\mathrm{GRECO}} + \ln\mathcal{L}_{\mathrm{GFU}}.  
 \end{equation}
 The TS is evaluated by comparing the signal hypothesis of this likelihood to the background hypothesis ($n_\mathrm{s}=0$) along with the spatial penalty from the GW probabilities. The pixel which maximizes this TS is chosen as the best-fit pixel, with the corresponding values for GFU and GRECO $n_\mathrm{s}$ as the best-fit values. Similar to the cascades analysis, the TS is derived for background, based on scrambled data, and is compared to pseudo experiments with neutrino injection to obtain the sensitivities. The background TS distributions are also used to obtain the p-values of the unblinded data and further constrain the flux upper limits from this analysis.
 \section{Results}
 \begin{figure}[h!]
\centering
\includegraphics[width=0.49\linewidth]{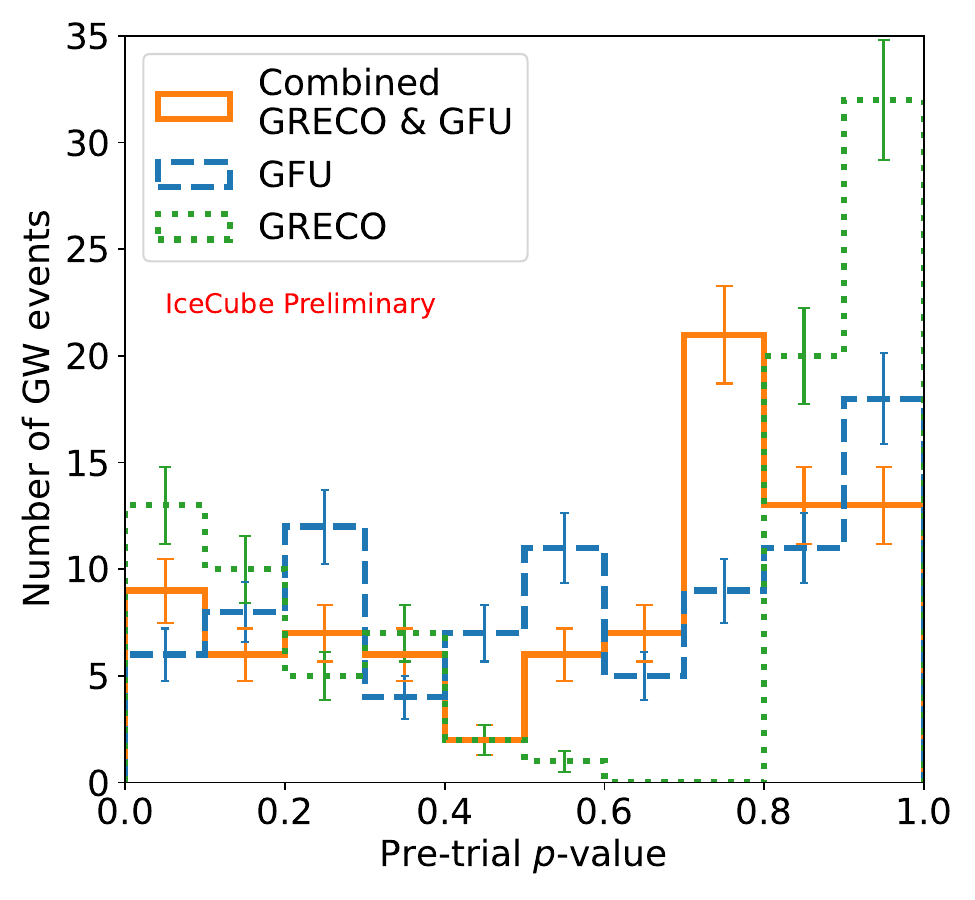}
\includegraphics[width=0.49\linewidth]{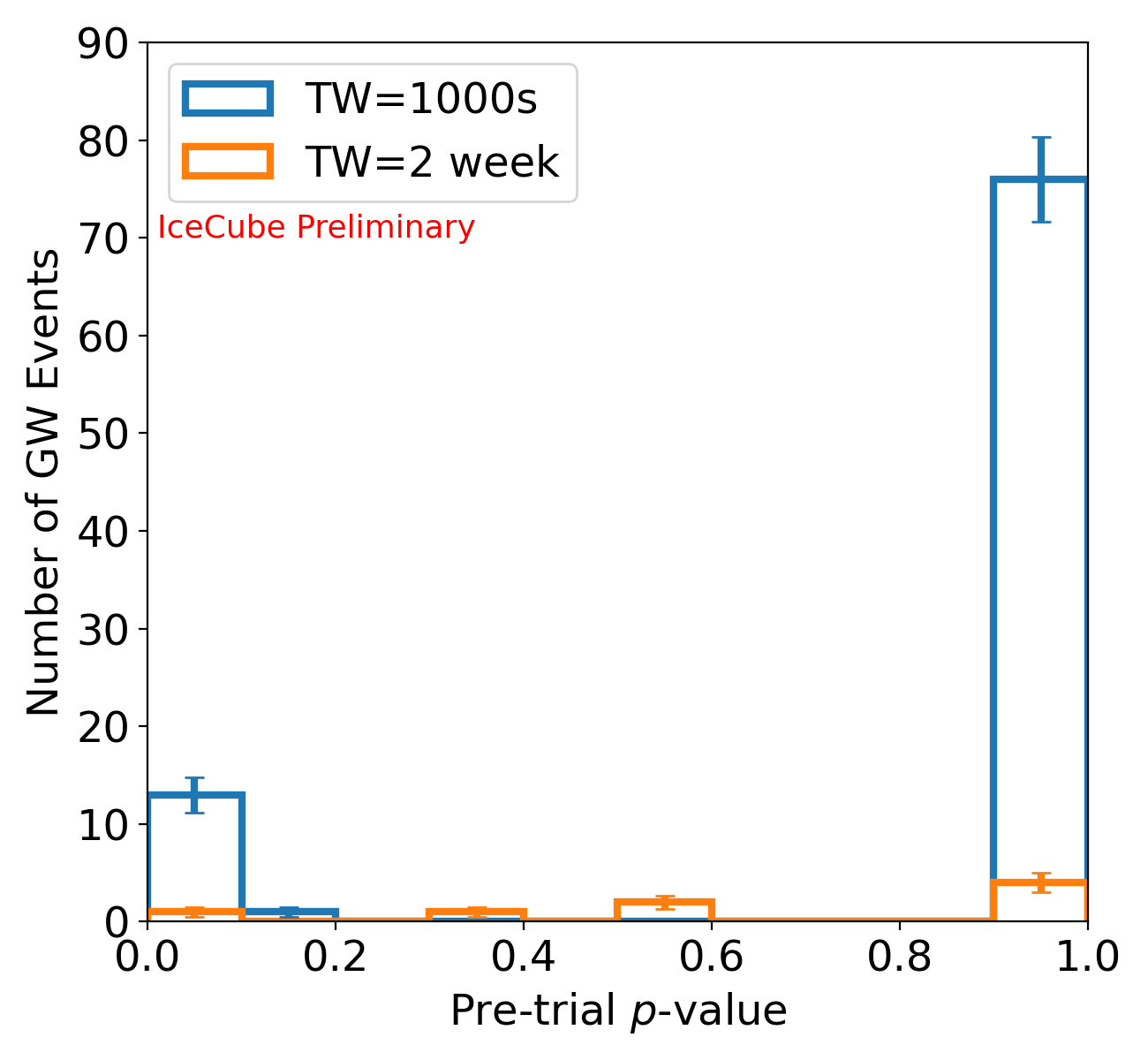}
\caption{Pre-trial p-value histograms of observed data. Left: The combined GRECO \& GFU analysis (orange) compared to the independently unblinded follow up with the GRECO Astronomy dataset (blue) and the GFU dataset (green), all for a 1000 second time window. Right: The observed p-values for the follow up using the cascade dataset. Both the 1000 seconds distribution (blue) and the 2-week distribution (in orange, only consisting of 7 NS containing GW events) are shown.}\label{fig1}
\end{figure}

\begin{figure}[h!]
\centering
\includegraphics[width=0.6\linewidth]{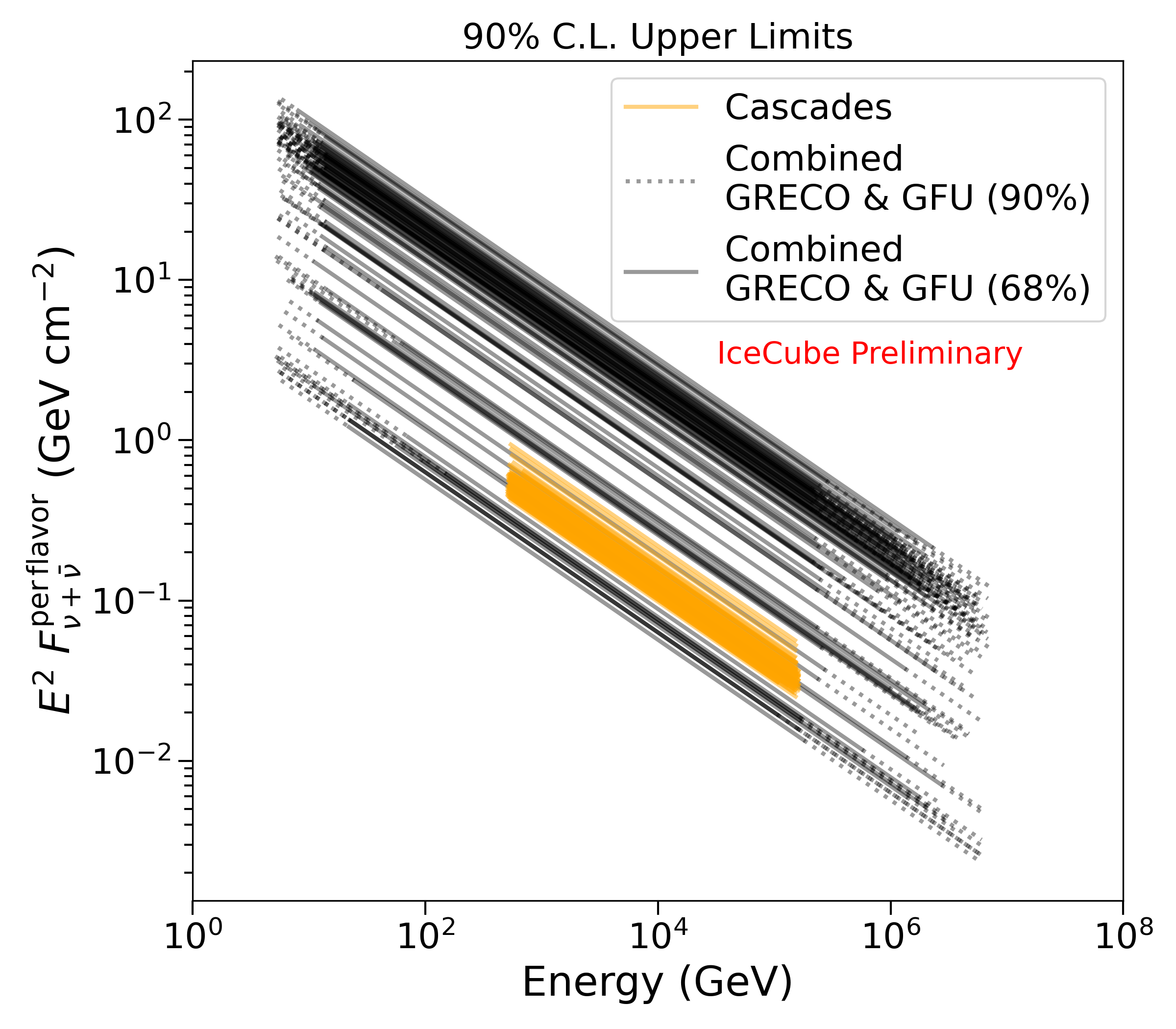}
\caption{Flux upper limits derived from the combined analysis (black) and the cascades analysis (orange) for a spectral index 2.5. The sensitive energy range (90\% CL and 68\% CL for the combined analysis and 90\% CL for the cascades analysis) across the declinations covered by the 90\% probablity region of the GW event of each analysis is highlighted.}\label{fig2}
\end{figure}
\begin{figure}[h]
\centering
\includegraphics[width=0.69\linewidth]{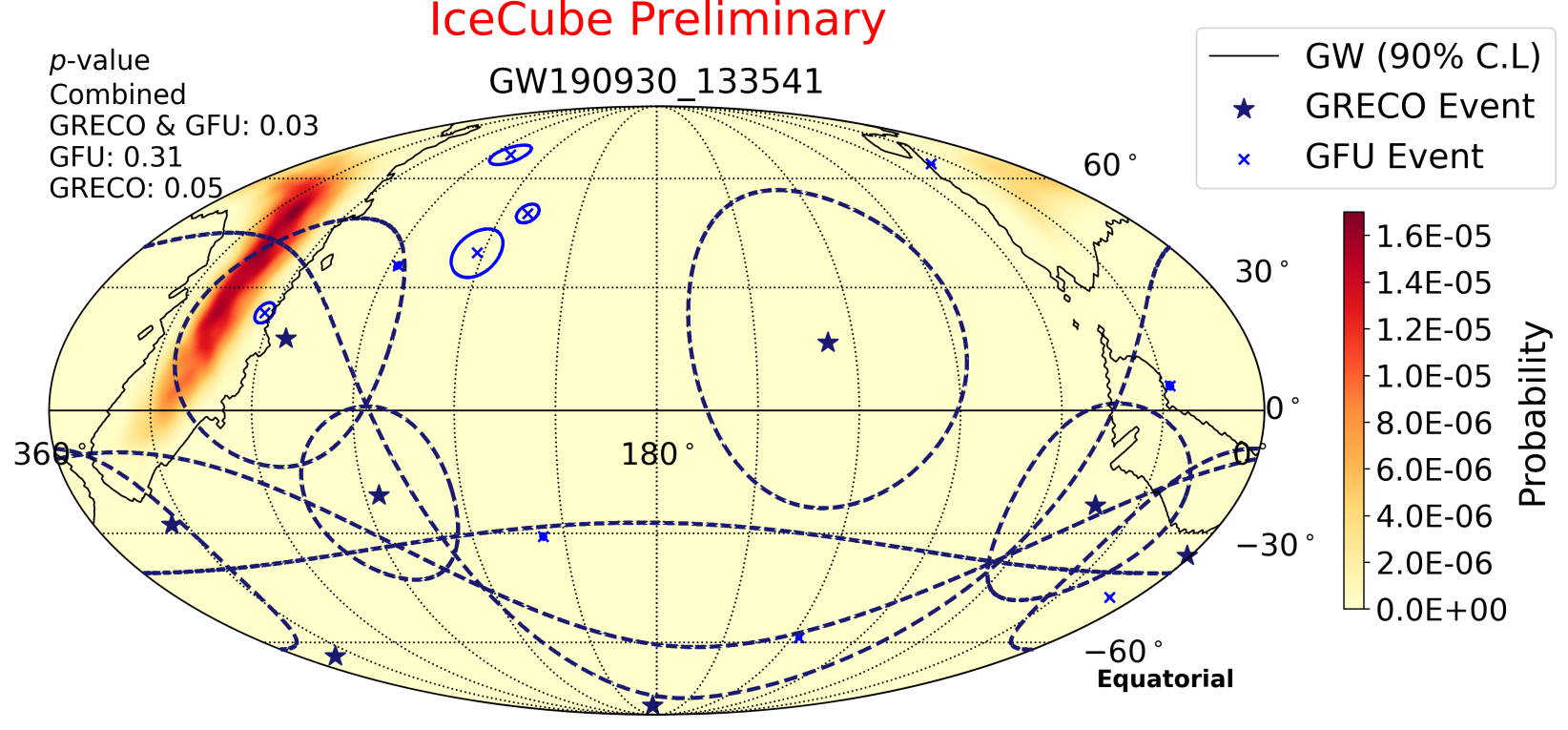}
\includegraphics[width=0.69\linewidth]{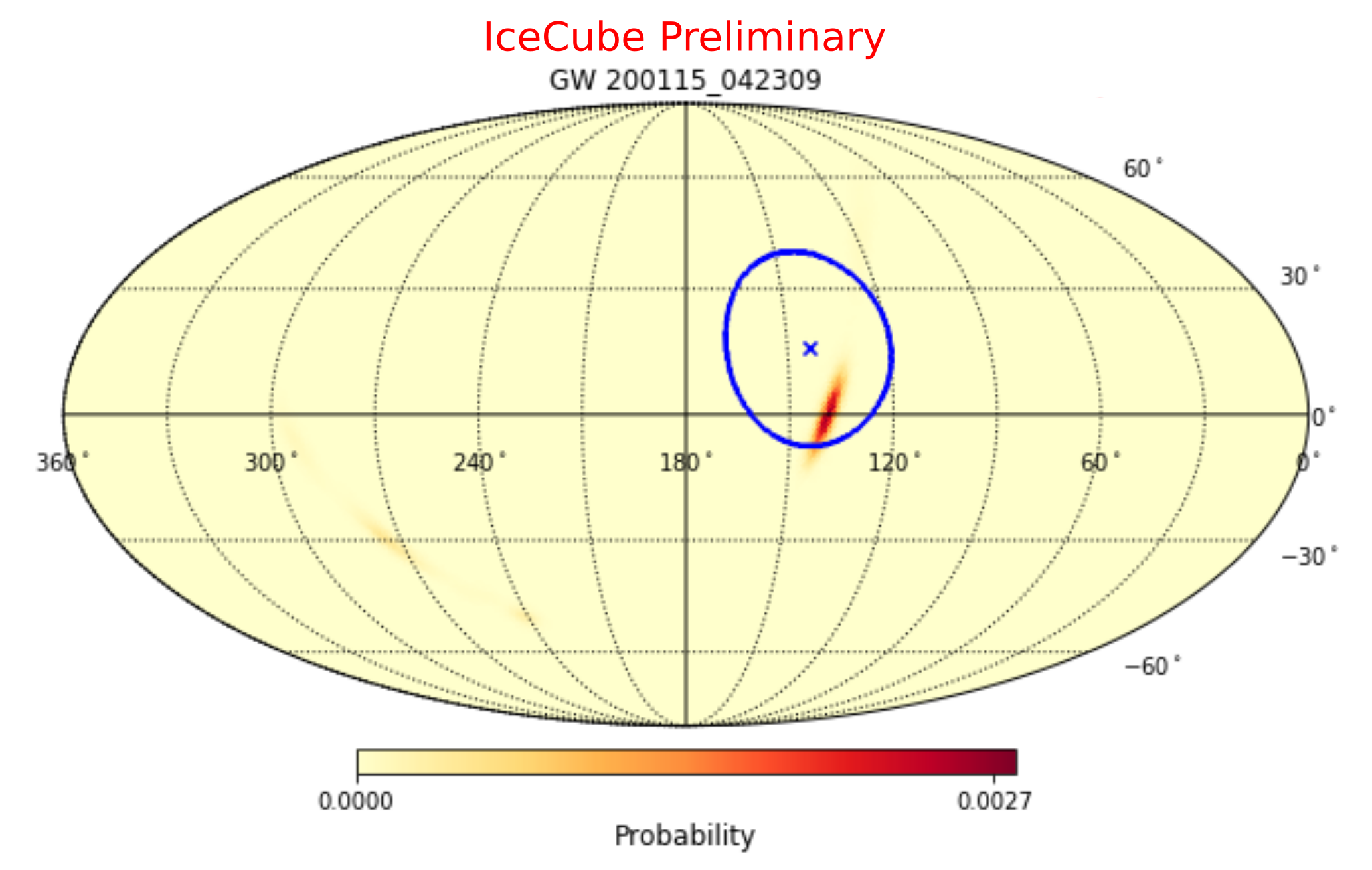}
\caption{Skymaps of events with lowest pre-trial p-value in the combined GRECO \& GFU analysis (top) and the cascades analysis (bottom). The colormap shows the GW probabilties on the sky for each event, obtained from the map released by LVK. The markers and the circles represent neutrinos along with their angular uncertainties, within the 1000 second time window. GFU events (top, solid blue) have smaller uncertainties since they are tracks of higher energies and therefore have better resolution. Cascade events (bottom, solid blue) and GRECO events (top, dashed navy blue) have larger angular uncertainties on the sky.}\label{fig3}
\end{figure}

We find no significant emission of neutrinos correlated to GW events from either analysis. The lowest pre-trial p-value for the combined analysis is 0.03 for the event GW190930\_133541. This event had a pre-trial p-value of 0.31 (0.05) in the independent GFU (GRECO) analysis. In the analysis with the cascades dataset, GW200115\_042309 was the most significant source in the 1000\,s time window with a pre-trial p-value of 0.002 while GW200115\_042309 was the most significant source in the 2 week test with a pre-trial p-value of 0.09.
Figure~\ref{fig1} shows the distribution of p-values obtained from both analyses and figure~\ref{fig2} shows the 90\% confidence level (CL) flux upper limits. Skymaps of the events with the lowest pre-trial p-values obtained with both analyses reported here are shown in figure~\ref{fig3} and a summary of the results for these events are shown in table~\ref{tab1}.

\begin{table}[b!]
\centering
\begin{tabular}{lccccccccccc}
\hline
Analysis  &	GW event & pre-trial & 90\% CL Flux UL\\
& & p-value  & $\mathit{E^2F^{\text{per flavor}}_{\nu+\bar{\nu}}}\mathrm{@1\,TeV}\ [\frac{\mathrm{GeV}}{\mathrm{cm}^2}]$\\
\hline
Cascades 1000s & 200115\_042309  & 0.002  & 0.60 \\
\hline
Cascades 2 week & 200115\_042309  & 0.09  & 1.60 \\
\hline
Combined GRECO \& GFU & 190930\_133541 & 0.03 & 0.22\\
\hline
\end{tabular}
\caption{Summary of results from the cascades analysis and the combined GRECO \& GFU analysis for the events with lowest p-value in both analyses. The flux upper limits assume a spectral shape $\propto E^{-2.5}$.}\label{tab1}
\end{table}

 We performed a binomial test for both analyses to search for a population of GW events producing an excess of neutrinos. This was not found to be significant in both analyses, with a final binomial probability of 0.017 (0.16) in the cascades analysis (combined analysis) corresponding to a population of 3 (89) GW events. After trials correction, where we compare the observed binomial probability to a distribution derived from background, these p-values become 0.07 and 0.56 for the cascades and the combined analyses respectively.

\begin{figure}[h]
\centering
\includegraphics[width=0.6\linewidth]{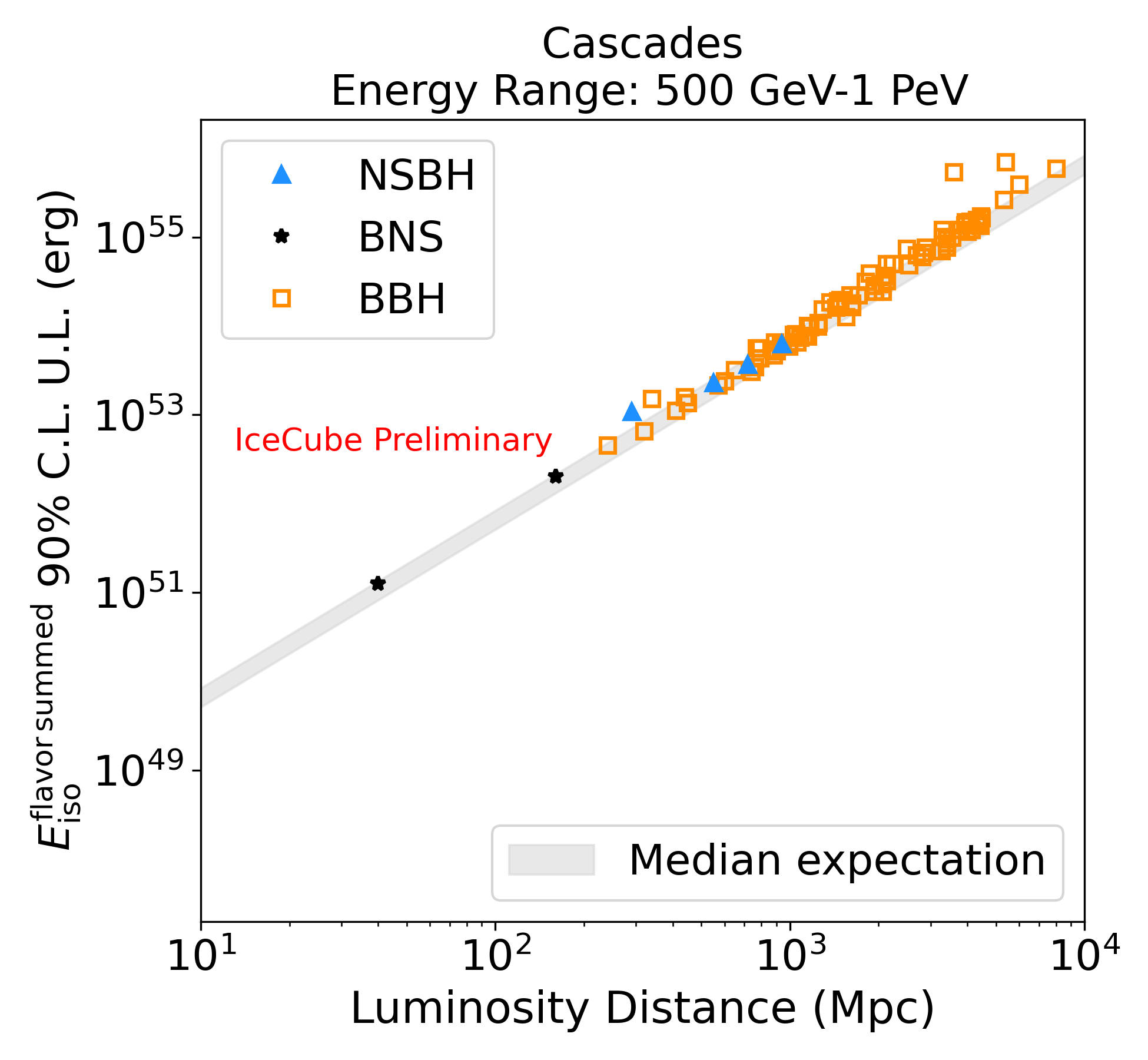}
\caption{Upper limits to the isotropic equivalent energy from the followup search using the cascades dataset. The upper limits are derived under the assumption of a neutrino flux following a spectral index of 2.5. BBH, NSBH, and BNS events are shown separately for 90 confident GW events from O1 to O3. The grey band shows the median expectation derived from sensitivity of the analysis, and is seen to follow the relation $E_{\mathrm{iso}}\propto r^2$.}\label{fig4}
\end{figure}

Upper limits to the isotropic equivalent energy emitted in neutrinos of all flavor were calculated with the cascades analysis. These limits assume that neutrinos are emitted with a spectral index of 2.5 within the energy range of 500\,GeV to 1\,PeV. These upper limits are shown in figure~\ref{fig4}. The figure also shows the median expectation, estimated from the sensitivity range of the cascade analysis to GW events from all declinations.

\section{Discussion and Conclusion}\label{sec3}
\begin{justify}
We report on the results of two novel analyses: the first search for cascade neutrinos from GW events and the first joint search of low and high energy neutrinos from GW events. Although no significant detection was observed in either analysis, the methodology for these two analyses set the groundwork for future searches. LVK sends public alerts of GW events with low latency, and IceCube currently searches for tracks that are spatially and temporally correlated to them. An introduction of cascade events to this real time stream can provide with additional sensitivity to such searches, especially in the southern sky. . The combined GRECO \& GFU analysis presented here provides a proof of concept for performing such joint searches in the future. Such searches can also be performed in realtime, providing added sensitivity. The combination of two datasets need not be restricted to low and high energy events alone. Analyses combining the strength of tracks and cascades for searching for multi-messenger emission can also be performed in the future.
\end{justify}
\bibliographystyle{ICRC}
\bibliography{references}

%

\clearpage

\input{authorlist_IceCube.tex}

\end{document}

%% file: ICRCdetails.tex
\ConferenceLogo{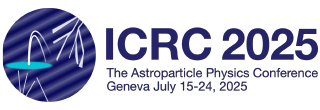}

\FullConference{39th International Cosmic Ray Conference (ICRC2025)\\
 15–24 July 2025\\
Geneva, Switzerland\\}

%% file: authorlist_IceCube.tex
\section*{Full Author List: IceCube Collaboration}

\scriptsize
\noindent
R. Abbasi$^{16}$,
M. Ackermann$^{63}$,
J. Adams$^{17}$,
S. K. Agarwalla$^{39,\: {\rm a}}$,
J. A. Aguilar$^{10}$,
M. Ahlers$^{21}$,
J.M. Alameddine$^{22}$,
S. Ali$^{35}$,
N. M. Amin$^{43}$,
K. Andeen$^{41}$
% ...
\\
\\
% ...
$^{10}$ Universit{\'e} Libre de Bruxelles, Science Faculty CP230, B-1050 Brussels, Belgium \\
% ...
$^{16}$ Department of Physics, Loyola University Chicago, Chicago, IL 60660, USA \\
$^{17}$ Dept. of Physics and Astronomy, University of Canterbury, Private Bag 4800, Christchurch, New Zealand \\
% ...
$^{21}$ Niels Bohr Institute, University of Copenhagen, DK-2100 Copenhagen, Denmark \\
$^{22}$ Dept. of Physics, TU Dortmund University, D-44221 Dortmund, Germany \\
% ...
$^{35}$ Dept. of Physics and Astronomy, University of Kansas, Lawrence, KS 66045, USA \\
% ...
$^{39}$ Dept. of Physics and Wisconsin IceCube Particle Astrophysics Center, University of Wisconsin{\textemdash}Madison, Madison, WI 53706, USA \\
% ...
$^{41}$ Department of Physics, Marquette University, Milwaukee, WI 53201, USA \\
% ...
$^{43}$ Bartol Research Institute and Dept. of Physics and Astronomy, University of Delaware, Newark, DE 19716, USA \\
% ...
$^{63}$ Deutsches Elektronen-Synchrotron DESY, Platanenallee 6, D-15738 Zeuthen, Germany \\
$^{\rm a}$ also at Institute of Physics, Sachivalaya Marg, Sainik School Post, Bhubaneswar 751005, India \\
% ...

\subsection*{Acknowledgments}

\noindent
The authors gratefully acknowledge the support from the following agencies and institutions:
USA {\textendash} U.S. National Science Foundation-Office of Polar Programs,
U.S. National Science Foundation-Physics Division,
U.S. National Science Foundation-EPSCoR,
U.S. National Science Foundation-Office of Advanced Cyberinfrastructure,
Wisconsin Alumni Research Foundation,
Center for High Throughput Computing (CHTC) at the University of Wisconsin{\textendash}Madison,
Open Science Grid (OSG),
Partnership to Advance Throughput Computing (PATh),
Advanced Cyberinfrastructure Coordination Ecosystem: Services {\&} Support (ACCESS),
Frontera and Ranch computing project at the Texas Advanced Computing Center,
U.S. Department of Energy-National Energy Research Scientific Computing Center,
Particle astrophysics research computing center at the University of Maryland,
Institute for Cyber-Enabled Research at Michigan State University,
Astroparticle physics computational facility at Marquette University,
NVIDIA Corporation,
and Google Cloud Platform;
Belgium {\textendash} Funds for Scientific Research (FRS-FNRS and FWO),
FWO Odysseus and Big Science programmes,
and Belgian Federal Science Policy Office (Belspo);
Germany {\textendash} Bundesministerium f{\"u}r Forschung, Technologie und Raumfahrt (BMFTR),
Deutsche Forschungsgemeinschaft (DFG),
Helmholtz Alliance for Astroparticle Physics (HAP),
Initiative and Networking Fund of the Helmholtz Association,
Deutsches Elektronen Synchrotron (DESY),
and High Performance Computing cluster of the RWTH Aachen;
Sweden {\textendash} Swedish Research Council,
Swedish Polar Research Secretariat,
Swedish National Infrastructure for Computing (SNIC),
and Knut and Alice Wallenberg Foundation;
European Union {\textendash} EGI Advanced Computing for research;
Australia {\textendash} Australian Research Council;
Canada {\textendash} Natural Sciences and Engineering Research Council of Canada,
Calcul Qu{\'e}bec, Compute Ontario, Canada Foundation for Innovation, WestGrid, and Digital Research Alliance of Canada;
Denmark {\textendash} Villum Fonden, Carlsberg Foundation, and European Commission;
New Zealand {\textendash} Marsden Fund;
Japan {\textendash} Japan Society for Promotion of Science (JSPS)
and Institute for Global Prominent Research (IGPR) of Chiba University;
Korea {\textendash} National Research Foundation of Korea (NRF);
Switzerland {\textendash} Swiss National Science Foundation (SNSF).